\documentclass[conference]{styles/sig-alternate}

\usepackage[
		   pdftex,
           pdfpagelabels,
           plainpages=false,
           bookmarks=true,
           hypertexnames=false,
           pdffitwindow=false,
           colorlinks=true,
           linkcolor=blue,
           citecolor=blue,
           filecolor=black,
           urlcolor=blue
           ]{hyperref}
\usepackage[T1]{fontenc}
\usepackage{flushend}
\usepackage{xspace,epsfig}
\usepackage{graphicx}
\usepackage{times,latexsym,amsfonts,amssymb,amsmath}
\usepackage{subfigure}
\usepackage{array}
\usepackage{comment,url}
\usepackage{multirow}
\usepackage{caption}
\DeclareCaptionType{copyrightbox}
\usepackage{enumitem}
\usepackage{amsmath}
\usepackage{epstopdf}
\usepackage{xspace,epsfig}
\usepackage[official]{eurosym}
\usepackage{graphicx,times,latexsym,amsfonts,amssymb,amsmath}
\usepackage{subfigure}
\usepackage{comment,url, xcolor}
\usepackage{multirow}
\usepackage{caption}
\usepackage{enumitem}
\usepackage{amsmath}
\usepackage{ragged2e}

\usepackage[numbers,sort&compress]{natbib}
\usepackage{microtype}

\newcommand{\youtube}{{YouTube}\xspace}
\newcommand{\dailymotion}{{Dailymotion}\xspace}

\newcommand{\vimeo}{Vimeo\xspace}

\newcommand{\myvideode}{{Myvideo.de}\xspace}
\newcommand{\tvuol}{{TV UOL}\xspace}

\newcommand{\bi}{\begin{itemize}}
\newcommand{\ei}{\end{itemize}}
\newcommand {\beq}{\begin{equation}}
\newcommand {\eeq}{\end{equation}}
\newcommand {\be}{\begin{enumerate}}
\newcommand {\ee}{\end{enumerate}}

\newcommand{\TONE}{\textit{YT-1}\xspace}
\newcommand{\TTWO}{\textit{YT-2}\xspace}

\newcommand{\specialcell}[2][c]{%
  \begin{tabular}[#1]{@{}c@{}}#2\end{tabular}}

\begin{document}
\begin{sloppypar}

\title{Understanding the Detection of View Fraud \\ in Video Content Portals}

\newcommand{\superscript}[1]{\ensuremath{^{\textrm{#1}}}}
\def\sharedaffiliation{\end{tabular}\newline\begin{tabular}{c}}

\def\wu{\superscript{*}}
\def\wg{\superscript{\dag}}
\def\wgg{\superscript{\ddag}}
\def\dm{\superscript{\S}}
\newcommand\blfootnote[1]{%
  \begingroup
  \renewcommand\thefootnote{}\footnote{#1}%
  \addtocounter{footnote}{-1}%
  \endgroup
}

\numberofauthors{6}
\author{
  Miriam Marciel\wg\wu, Ruben Cuevas\wu, Albert Banchs\wgg\wu,
  Roberto Gonzalez\wg, Stefano Traverso\dm,\\
   Mohamed Ahmed\wg, Arturo Azcorra\wgg\wu
  \vspace{-.15cm}
  \sharedaffiliation
    \affaddr{{\wg}NEC Labs Europe},  
    \affaddr{{\wu}Universidad Carlos III de Madrid}, 
    \affaddr{{\wgg}IMDEA Networks Institute},
    \affaddr{{\dm}Politecnico di Torino} \\
    \affaddr{\{miriam.marciel, roberto.gonzalez, mohamed.ahmed\}@neclab.eu,
      \{rcuevas, banchs, azcorra\}@it.uc3m.es,} 
    \\
    \affaddr{stefano.traverso@polito.it}
}
\maketitle

\begin{abstract}

While substantial effort has been devoted to understand fraudulent
activity in traditional online advertising (search and banner), more
recent forms such as video ads have received little attention. The
understanding and identification of fraudulent activity (i.e., fake
views) in video ads for advertisers, is complicated as they
rely exclusively on the detection mechanisms deployed by video hosting
portals. In this context, the development of independent tools able to
monitor and audit the fidelity of these systems are missing today and
needed by both industry and regulators.

In this paper we present a first set of tools to serve this
purpose. Using our tools, we evaluate the performance of the audit
 systems of five major online video portals. Our results
reveal that \youtube's detection system significantly outperforms all
the others. Despite this, a systematic evaluation indicates that it
may still be susceptible to simple attacks. Furthermore, we find that
\youtube penalizes its videos' public and monetized view counters
differently, the former being more aggressive. This
means that views identified as fake and discounted from the public
view counter are still monetized.  We speculate that even though
\youtube's policy puts in lots of effort to compensate users after an
attack is discovered, this practice places the burden of the risk on
the advertisers, who pay to get their ads displayed.

\end{abstract}

\begin{keywords}
Fraud, fake views, YouTube, active probing, advertising
\end{keywords}

   
\section{Introduction}%

The Interactive Advertisement Bureau (IAB) reported that online advertising generated revenue of \$49B in 2014, in the U.S. alone. This figure corresponds to a 15.6\% increase in revenue with respect to 2013~\cite{IAB}. Of particular interest to this work is video advertising. A recent survey indicates that in 2013, 93\% of online marketers used video to advertise their products, and of these, 65\% used \youtube specifically to deliver the content~\cite{emarketer:video:2013}. Online video advertising is estimated to have generated \$3.3B in 2014, in the U.S. alone; approximately 7\% of the total revenue generated by online advertising~\cite{IAB}. This figure is estimated to have risen from \$2.2B in 2012, and is expected to grow to \$8B by 2016~\cite{IAB,emarketer:value-of-ads:2013}.

Given such revenues, it is no surprise that online advertising attracts fraud. Recent studies have estimated that 15-30\% of ad impressions to be fraudulent~\cite{Luttrell:buyerside:2013,wsj:online-ads-fraud}, and in some portals this number may be as high as 75\%~\cite{neal2015quantifying}. This is estimated to lead to losses in the order of billions of dollars a year for advertisers \cite{click_fraud_cost}. With respect to online video ads, the media and online advertising industry both report that fraud is endemic~\cite{video_ads_attack_1,video_ads_attack_2,proRussiaVideoBots}. The U.S. based marketing representative body, the Association of National Advertisers (ANA), reported in 2014 that on average 23\% of video ad views across different studies were fraudulent~\cite{botBaseLine:14}.

In contrast to ``click fraud'' in search and display advertising (cf. ~\cite{Dave:2013:VCC:2508859.2516688,Metwally:2007:DDC:1242572.1242606,Stone-Gross:2011:UFA:2068816.2068843,iab_antifraud_wg}), fraud in online video advertising has received comparatively little attention~\cite{chen2015analysis}. Typically, the goal of click fraud is to inflate user activity counters at a particular target, such as a webpage.  Online video ads however offer new motivations, attack paths, and revenue streams. First, the status and earning from uploading popular online videos~\cite{wt_ty_millionaires_15} commonly attracts fraudulent activity~\cite{Tubrosa}, which has triggered online video portals to start auditing their systems~\cite{yt_fakeviews_policy}. For example, it was reported in 2012 that~\youtube removed more than 2B suspected ``fraudulent'' views from accounts associated with the music industry~\cite{hoffberger:youtube-strip-2012}. Second, in contrast to search and banner advertising, where advertisers can collect partial information on their users from clickbacks. Online videos advertisers must delegate the detection and auditing of fraud to the portals that host their content, and rely on the high-level statistics they offer.
Finally, while search and banner ads are sold at either \emph{Cost-per-Impression (CPI)}, or \emph{Cost-per-Click (CPC)}, video ads are typically sold at \emph{Cost-per-View (CPV)}~\cite{cpv}, which are on average more expensive (sold in sets of 1000, and referred to as Cost Per Mille (CPM))~\cite{tubeMogulPlaybook}.

The common attack in online video ad fraud is to inflate the view counters of videos using botnets~\cite{supremetrafficbot}, or crowd sourced users~\cite{AIB:fraudTaxonomy}. In fact, it is easy to find paid services that generate tens of thousands of views to videos hosted on popular portals (e.g. \youtube, \dailymotion and \vimeo) at a low price~\cite{viewbros,qqtube,buildmyviews}. If the goal of the attacker is simply to increase the popularity and visibility, of their videos, then this is enough. If however, the goal is to generate revenue, then the attacker attempts to have ads served to their fake viewers, and collects a share of the revenue.

In response to the scale of the video-ad fraud, the media and online advertisers have consistently publicized the need for more effective anti-fraud solutions~\cite{bloomberg_fakeads_rotting,guardian_youtube_age,ft_2,isba}. The IAB has recently formed a working group to address the problem~\cite{iab_antifraud_wg}, and has so far published a white-paper report on anti-fraud principles, and proposed a reference taxonomy~\cite{AIB:fraudTaxonomy}. Finally,  some online video portals have acted to forestall damages by strengthening their view auditing systems and publicizing their activity~\cite{youtube_reaction,googleWarOnBots}.

Despite these initial steps, today we lack the tools, methods, and standards to independently understand, audit, and monitor the function and performance of the fraud detection mechanism deployed by popular online portals. This is reflected in the IAB working group white-paper which states that ``\textit{[Supply sources] are challenged by a lack of consistent and independently measurable principles on how they each should identify and expunge fraudulent traffic}''~\cite{AIB:fraudTaxonomy}. 

The main contribution of this paper is a novel measurement methodology to aid in filling this gap. Employing a modular active probe, we evaluate the performance of the fraud detection mechanism (for public and/or monetized views) of 5 online video portals, namely \youtube, \dailymotion, \vimeo, \myvideode, and \tvuol.

Finding that \youtube is the only portal deploying a sufficiently discriminative view audit system, we deepen our analysis to study some of its key parameters. We focus on parameters that are directly accessible to users, and are reported to be manipulated by video view-inflation bots in the wild \cite{Tubrosa,YoutubeBotViews,YouTubeBotViewsProxies}.

We study the impact of manipulating the behavior of an IP address, such as varying the number of videos visited per day, the views per video, and the duration per view. We then look at the impact of changing the browser-profile of viewers, such as whether or not cookies are enabled, and the impact of mixing viewer activity in NATed traffic.


\noindent Our main findings can be summarized as follows:\\
\noindent \textbf{(1)}  Of the 5 portals listed, \youtube is the only portal to deploy a significantly discriminative view audit system for the public view counters. All other portals do not sufficiently discount their view counters, even under the simplest fake views generation configurations.

\noindent \textbf{(2)} A deeper analysis reveals that the detection mechanisms of \youtube's public view counter are susceptible to simple fake views generation strategies such as; using multiple values in the HTTP connection attributes (e.g., User-Agent or Referrer), distributing views across multiple IP addresses, or routing views through NATs.

\noindent \textbf{(3)} We find a consistent and significant discrepancy between the counter values reported for the same content by the public, and monetized view counters in \youtube. We find that the monetized view counters count at least 75\% more fake views than public view counters.

\noindent \textbf{Organization of the paper\\}
The rest of the paper is organized as follows. Sec.~\ref{sec:background} presents the background on the business models and statistic reporting tools for the five online video portals considered in this study. In Sec.~\ref{sec:software_description} we present the measurement tools and the performance metrics used in this study. Sec.~\ref{sec:comparison_portals} evaluates the performance of the view audit systems of the different online portals. Sec.~\ref{sec:youtube} and Sec.~\ref{sec:youtube_monetization} present more detailed analysis of how \youtube's audit systems discount the counters for the public and the monetized view counters, respectively. Finally, Sec.~\ref{sec:relatedwork} discusses the related work, Sec.~\ref{sec:ethical} discusses the ethical considerations and feedback received from the industry and Sec.~\ref{sec:conclusions} concludes the paper.

\section{Background}
\label{sec:background}

In this work, we focus on user-generated video portals, the most widely used and, therefore the most susceptible to video advertising fraud. Table~\ref{tab:market_share} summarizes the online-video market shares of the portals considered in this study. Since \youtube is reported by all sources to be the largest portal, it will serve as the reference portal in our study.

\begin{table}[t!]
  \scriptsize
  \centering
  \begin{tabular}{c||c|c |c|c|c}
    \textbf{Source}	& \youtube & \vimeo & \dailymotion & \myvideode & \tvuol \\
    \hline
    \hline
    SYSOMOS~\cite{sysomos}	& 81.9\% 	& 8.8\% 	&4\% 	& - 	& - \\
    DATANYZE~\cite{datanyze}	& 65.1\% 	& 11.1\% 	& 0.6\% 	& - & - \\
    NIELSEN~\cite{nielsen} 	& 84.2\% 	& - & 1.16\% 	& - & - \\
    Statista~\cite{statista} & 73.6\% 	& 0.9\% 	& 1.6\% 	& - & - \\
    \hline 
    \shortstack{Alexa}	& 3 & 145 & 84 & \begin{tabular}{@{}c@{}} 3236 \\(DE: 153)\end{tabular} & \begin{tabular}{@{}c@{}}101 \\(BR: 5)\end{tabular} \\
    \hline 
    \shortstack{\begin{tabular}{@{}c@{}}Views/day\\ (x1M)~\protect\cite{wikipedia}\end{tabular}} & 1200 & 1 & 60 & 7 & 6
  \end{tabular}
  \caption{Market share and rank of the portals studied from different
    public sources. 
  }
  \label{tab:market_share}
\end{table}

User-generated video portals typically monetize the content uploaded by their users through advertising. \youtube, \dailymotion, \myvideode, \tvuol all deliver ads on the videos streamed to their viewers. \youtube directly incentivises its users by sharing with them, the ad revenues generated by views to the videos they upload and explicitly enroll into its monetization programme.  \dailymotion instead incentivises third party web masters, by sharing with them ad revenue generated from views to videos embedded on their sites.\footnote{Web masters can embed any video available on \dailymotion in their website.} In contrast, \vimeo runs a subscription based model. Users subscribed to its `Plus' account are able to monetize their uploads by using the ``Tip Jar'' service, that enables other viewers to tip to the uploader. Moreover uploaders subscribed to its `Pro' service may use a ``Pay-To-View'' service in which viewers pay to watch.
Finally, while \myvideode, and \tvuol show ads in videos, to the best of our knowledge, they do not share ad revenue with their users. 

Under these revenue models, malicious users are incentived to inflate their view counters because revenue is divided based on view counts, as in the case of \youtube and \dailymotion. 
However, as mentioned the goal of user view inflation is not just limited to defrauding revenue from ad systems. There are numerous documented cases showing that users can, and do trade on just the popularity of their uploads, cf.~\cite{youtube_music_fakes}.

To help their uploaders understand how viewers interact with their content, video portals report various statistics to them.\\ 
\noindent \textbf{\youtube} provides two main sources of data on user activity and counted views; public statistics (public view counter, number of comments, likes, dislikes, number of subscribers) that are available on the video page, and private statistics (referred to as \youtube Analytics) that include the number of counted and monetized views, and are only available to the video uploader. 

\youtube Analytics provides detailed statistics, including; the number of video views grouped by day, country, viewer age, gender, or the playback location (if video is embedded in third party websites). Uploaders are also given summary reports on their channel subscribers, including their likes and dislikes, comments, etc. Finally, these statistics are updated daily \cite{views_report}, and based on our experiments, \youtube Analytics counters include only the validated views.

\youtube provides separate statistics for counters on monetized content. To monetize their content, uploaders have to first create an AdSense account, and enroll their \youtube channel. Uploaders can then view monetization statistics in both their \youtube Analytics and the AdSense accounts. In this paper we use the monetization statistics from the \youtube Analytics service, which is claimed to provide an error of less than $\pm$ 2\% with respect to the actual number of monetized views~\cite{yt_estimated_earnings}.
In particular, the monetization statistics offered by \youtube Analytics are; $(i)$ the estimated number of monetized views, i.e., the number of views that see an associated video ad, $(ii)$ the estimated revenue based on the Cost per Mile (CPM), and $(iii)$ the total gross revenue the video generated. In order to enable uploaders to better target their contents, these metrics are available by country, date and type of ad. 
 
\noindent \textbf{\dailymotion} provides public view counts on each video page, and uploaders can access similar statistics to those offered by \youtube Analytics. For example the number of views filtered by country, and playback location, over a selected time window.  Web masters registered with the \dailymotion monetization service can access monetization statistics including the number of impressions, the estimated revenue, and CPM. However, these statistics are aggregated across all videos associated to a web master's account, and are not available for individual videos.

\noindent \textbf{\myvideode and \tvuol} provide public view counters only. This data can be accessed through the video page and via the uploader account.

\noindent \textbf{\vimeo} offers public statistics for each video including the number of views, likes, comments, as well as their weekly evolution. Vimeo's default account type reports to its users only the public statistics, whereas the Vimeo `Plus' and `Pro' accounts provide more detailed statistics~\cite{vimeo_statistics}, including; geographical information about the views, information about user comments, or likes for the video, etc.

In addition, video portals offer advertisers statistics on the performance of their video ads. For instance, \youtube uses Google AdWords (Google's advertising campaigns service) for this purpose. Among other statistics, Google Adwords provides information about the number of views charged to advertisers, as well as the videos where their ads were shown. These statistics are aggregated by day.

Since the statistics reported by portals are typically summaries, it is difficult for third parties to understand how they are generated. Therefore, this work helps to address this gap by proposing and testing view counter auditing tools and methodologies for online video portals.

\section{Measurement Tools, Performance Assessment Methodology  and Data Processing}
\label{sec:software_description}

In this section we present the methodology and tools developed to independently evaluate the effectiveness of the view audit mechanisms deployed by the online video portals listed in Table~\ref{tab:market_share}.

Given that we are not able to observe all the data collected by portals on their users, nor the logic of their audit systems, in this work, we simplify the problem by exploring only parameters and methods that are directly accessible to third parties (uploaders and viewers). Specifically, we explore the impact of the viewer behavior and the viewer IP address space.

\subsection{Active Measurement Tools}\label{subsec:bot}

To study the performance of the view audit systems deployed by the portals, we deploy active probes that auto generate views, under well defined constraints, and log the results of their activity. In addition, we utilize tailored web crawlers to collect the statistics provided by the different video portals, such as the numbers of counted and monetized views.

\noindent \textbf{Automatic Views Generation:} \label{subsec:bot} 
We implemented a Selenium~\cite{Selenium} based (modular) probe to simulate the actions of viewers on the different portals. The probe is able to load a given video page, and can be easily configured to perform certain viewer-like actions, such as, interacting with the objects in the page, or varying the duration of video views. The different configurable parameters of our tool are similar to those of some well-known malware, devoted to fraudulently viewing \youtube videos, \cite{Tubrosa,YoutubeBotViews,YouTubeBotViewsProxies} and are therefore representative of realistic attack configurations observed in the wild.
Table~\ref{tab:botparams} summarizes the list of available parameters and their default settings. 

\begin{table}[t!]
  \scriptsize
  \begin{tabular}{>{\RaggedLeft}m{0.2\linewidth}|| m{0.5\linewidth} | >{\RaggedRight}m{0.2\linewidth}}
   \textbf{Parameter} & \specialcell{\textbf{Description}} & \textbf{Default Value} \\
   \hline
   User-Agent & Set the User-Agent for a session (e.g., Firefox or Chrome).  & Linux/Firefox \\\hline
   Referrer & Set the referrer for a session. Options are Facebook, Twitter, \youtube Search (specific for the case of \youtube), and Direct Link. & Direct Link\\\hline
   Cookies & When enabled, all the views have the same cookies.  & Disabled \\\hline
   View duration & Duration of a video view (in seconds). Options are i) fixed time or ii) samples from an exponentially distributed random variable with mean the duration of the video. & Fixed ($40$ secs.) \\\hline
   Wait time between views & Vary the view inter-arrival time (seconds). Options are a Poisson process, or a constant. Zero indicates a burst. & Constant factor of the number of daily views \\
  \end{tabular}
  \caption{Description of the software probe parameters and their default values.}
  \label{tab:botparams}
   \vspace{-0.3cm}
\end{table}

{\bf \noindent Experiment Isolation}: In order to isolate the impact of the experiments on the portals, we limit the maximum number of views generated by the experiments, and the probes generate views to \underline{only} videos that we upload for the experiments. All experiments are repeated multiple times in order to make the statistics robust. To reduce the impact of background noise, such as real users stumbling upon the videos, we set the names and descriptions of all experiment videos to random hashes, and all external links to them are removed.  To get a baseline for the effectiveness of the method, we measure the scale of the background noise of our approach by uploading 209 videos to \youtube, which we find attract only 21 views in total from external users in a three month test period.

To conduct the experiments, we use $\sim$100 public IP addresses located in two different /24 prefixes  in Spain and Germany. Moreover,  we install transparent proxies (Squid~\cite{squid}) in 300 PlanetLab nodes  PlanetLab nodes~\cite{Planetlab} and use 70 of them in the experiments.
The proxies relay views generated by  probes coordinated from a centralized controller. Finally, through experimenting we determine that \youtube treats direct and transparently proxied requests equally.

\noindent \textbf{Fetching Statistics from Video Portals:}
To retrieve the statistics reported by each portal, we deploy portal-tailored web crawlers. These enable us to: $(i)$ collect the information from the video public view counters,  $(ii)$ login to the uploader
account and retrieve the number of counted views for the video, as well as the number of monetized views (if available).\footnote{In the case of Dailymotion, the crawler can also login in the web master account.} In particular, for Myvideo.de, TV UOL, and Vimeo, we retrieve information on the number of counted views, whereas for \youtube and \dailymotion we  also retrieve the reported statistics for the number of monetized views.

\subsection{Performance Analysis}
To measure the performance of the different portal view audit systems and compare them, we analyze their classification accuracy. We measure their accuracy in detecting fake views, and report the false negative rates. For some specific portals we also report their false positive rates.

\vspace{0.05cm}

\noindent{\bf False negative rate:}  a \emph{false negative} is a `fake view' that is misclassified and counted in the view counter (public or monetized).
To measure the false negative ratio of a portal, a probe generates views to given videos and retrieves the number of counted views from the statistics offered by the portal. The false negative rate ($R_{FN}$) for the given platform is defined as:
\vspace{0.1cm}
\begin{equation*}
    R_{FN} = \frac{\text{\# counted views}}{\text{\# `probe' generated views}}
\end{equation*}
\vspace{0.1cm}

\noindent{\bf False positive rate:} a \emph{false positive} is defined as a `real user' view that is labeled as fake by the view audit systems, and not counted in the view counters.
To measure the false positive ratio of a portal, we crowd-source real users to view experiment videos on the portals, and then retrieve the view statistics from the portal. To accurately count user views, we first embed the videos into webpages that we can monitor, and then count only views via the webpage. The false positive rate ($R_{FP}$) for the given platform is then defined as:
\vspace{0.1cm}
\begin{equation*}
    R_{FP} = 1 - \frac{\text{\# counted views}}{\text{\# `real-user' generated views}}
\end{equation*}
\vspace{0.1cm}

\noindent{\bf Data Processing:}
In carrying out these experiments, we found that the view audit systems of some video portals displayed temporally transient behaviors. In particular, we observed that some experiments showed peaks in the false negative ratio on a unique day. Moreover, YouTube would count more views in the first few days of the experiment, then later adjust to a lower stable false negative ratio for the rest of the experiment. Therefore, in order to identify the standard behavior of view audit systems of video portals, and remove the impact of transients, we compute for each experiment the daily false negative ratio and calculate the median across the days of the experiment. 
For simplicity, we refer to this metric as $R_{FN}$ in the rest of the paper. Finally, we take care of repeating all experiments numerous times to provide statistical
confidence and report average, min and max $R_{FN}$ across experiments.

\section{View Fraud Detection in Online Video Portals}
\label{sec:comparison_portals}

In this section, we investigate how views are counted by the different portals listed in Table~\ref{tab:market_share}. We first compare how the portals penalize views in their public view counters. We then look at how \youtube and \dailymotion, which share revenue with their users, penalize view counters for monetized content.

\begin{table}[t!]
\scriptsize
\centering
\begin{tabular}{c || c |c |c |c |c}
\textbf{Trace}     & \textbf{Period} & \textbf{Length}      & \textbf{\# IP addresses}  & \textbf{\# Views}         & \textbf{\# Videos} \cr\hline
\TONE\    & 01/03/13-30/04/13       & 2 months  & 28071         & 3.94M         & 1.37M\cr \hline
\TTWO\   & 01/05/13-30/11/13    & 7 months  & 16781         & 15.9M			& 3.95M\cr
\end{tabular}
\caption{Summary statistics of measurement traces containing \youtube video sessions.}
\label{tab:traces}
\vspace{-0.2cm}
\end{table}

\subsection{The Accuracy of public view counters}\label{sec:public_counters}
\noindent{\bf Rate of False Negatives:}
We start by looking at the rate of false negatives for public view counters. To do so, we set up a simple experiment whereby each probe, with default parameters, from a fixed IP address, varies only the number of views it generates per day,  to a given video, on a given portal. In particular, for each portal, we generate 100, 400 and 500 views per day, which corresponds to view inter-arrival times of 864, 216, and 172 sec.,
to targeted videos. Each experiment runs for eight days and is repeated three times using IP addresses from our prefixes in Germany and Spain. 

To understand whether the number of views that we generate corresponds to normal user behavior, we collect traces from a residential ISP, and log the YouTube sessions.
We replicate the methodology described in \cite{Finamore:2011:YEI:2068816.2068849}, and collect two independent datasets (from the residential network of an ISP) that contain millions of YouTube sessions. We refer to these datasets as $\TONE$ and $\TTWO$ and we summarize their main characteristics in Table~\ref{tab:traces}. Our traces indicate that no single IP addresses in $\TONE$ and $\TTWO$ performs more than 100 views per day to a single video.\footnote{Since \youtube is the most popular among the portals studied, we assume that the configured number of views per day represents an aggressive setup for all the portals.} Therefore these configurations (100, 400, 500 views per day) should correspond to an `aggressive' probe behavior, and we expect them to be detected easily. 

\begin{figure}[t]
  \centering \includegraphics[width=0.47\textwidth]{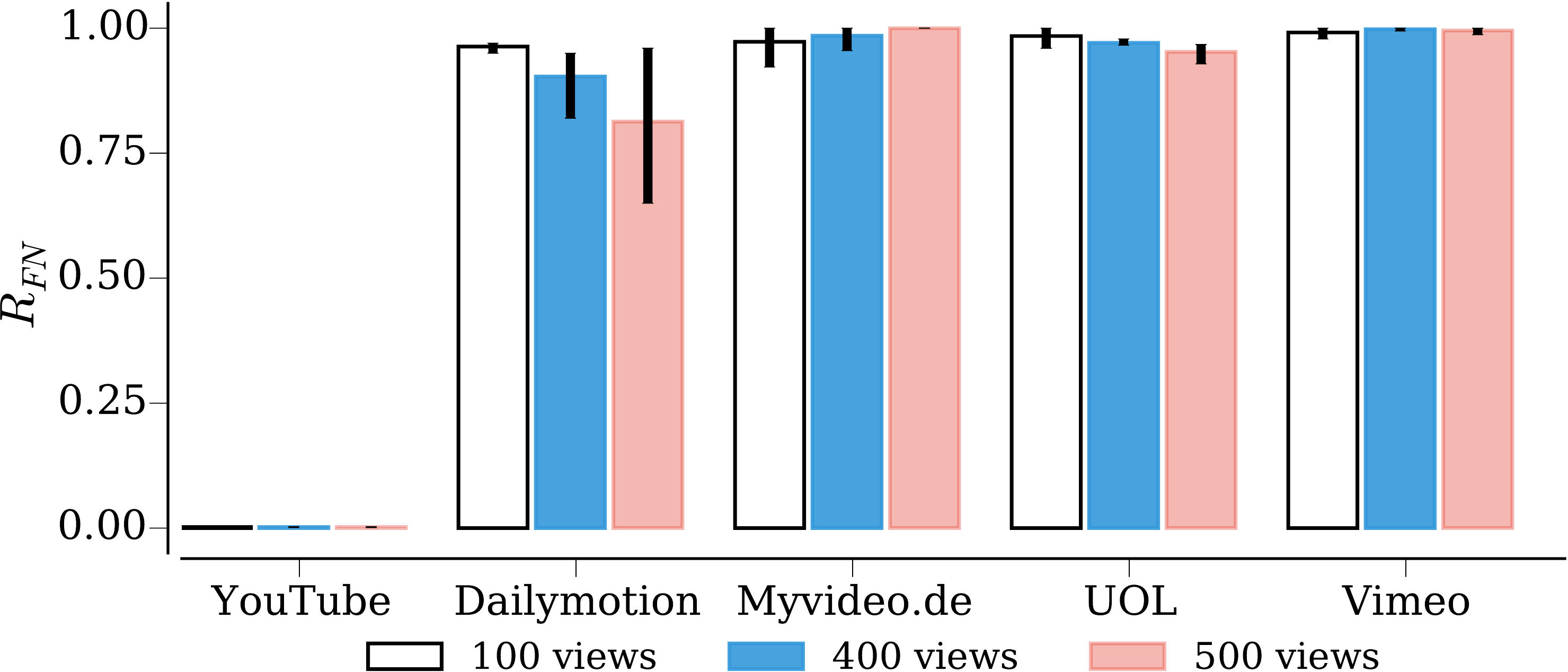}
  \caption{Comparison of the false negative ratio of the public view counter of the studied portals for different daily view rates.}
  \label{fig:comparison_portals}
  \vspace{-0.3cm}
\end{figure}

The results of this experiment are reported in Figure~\ref{fig:comparison_portals}.
The main bars report the average rate of false negatives for the three experiments, while the error bars report the max and min $R_{FN}$. The different colors correspond to the different daily view rates.

Our results indicate that \youtube, which penalizes \textbf{all} the views from the probes, operates the most discriminative view auditing system, and is significantly more effective than the other portals.  In contrast, \dailymotion counts as valid almost all the views when the daily rate is 100, and 93\% (85\%) when the rate is 400 (500) views per day. \myvideode, \tvuol and \vimeo deploy view audit systems that detect $<$ 5\% of the probe's fake views, even for the most aggressive configuration.

In summary, we observe that \youtube implements the most discriminative view audit system, and is able to easily detect `obviously' aggressive behaviors. Surprisingly, the systems deployed by all the other portals appear to be almost completely ineffective.  

\vspace{.2cm}
\noindent{\bf Rate of False Positives:}
To evaluate the rate of false positives, we embed videos hosted in the different portals into webpages we control, and record the number of real users accessing each page and watching each video, and the duration of each view. We compare the impact of sourcing users via social media and an online crowdsourcing platform.

In the case of social media, we recruit volunteers viewers by advertising for the experiment URLs on Facebook and Twitter. In the case of the crowdsourced users, we use a crowdsourcing platform to recruit paid viewers. Finally, since the results in Figure~\ref{fig:comparison_portals} indicate that only \youtube and \dailymotion are significantly discriminative in updating public view counters, we evaluate the rate of false positives only for these two portals.

The resulting false positive rates of the experiments are summarized in Table~\ref{tab:false_negatives_yt}. We find that the $R_{FP}$ is reasonably small for the two portals under both user sourcing approaches ($<12\%$). From this, we conclude that the view audit systems of \youtube and \dailymotion are fairly effective at identifying views generated by real users. However, we note that \dailymotion shows a larger $R_{FP}$ for the first experiment.

Finally, it is worth noting that the data provided by \youtube and \dailymotion in both experiments shows a  spatially localized distribution of viewer visits. 
For the social media experiment, most of the views come from Spain, whereas, for the experiment that uses crowdsourcing, most of the views come from India and Bangladesh.

\begin{table}[t]
\small
\begin{tabular}{c || c |c|c |c }
\textbf{Platform} & \textbf{Experiment} & \specialcell{\textbf{\# performed} \\ \textbf{real views}} &  \specialcell{\textbf{\# counted} \\ \textbf{views}}  &   $R_{FP}$ \\ \hline
\multirow{2}{*}{\youtube} & \specialcell{Social Media} & 330 & 322 & 2,4\% \\
& \specialcell{Crowdsourcing} & 599 & 537 & 10,3\% \\ \hline
\multirow{2}{*}{\dailymotion} & \specialcell{Social Media} & 325 & 290 & 10.9\% \\
& \specialcell{Crowdsourcing} & 587 & 515 & 12.2\% \\
\end{tabular}
\caption{False positive ratio for the social media and crowdsourced experiments for \youtube and \dailymotion.}
\label{tab:false_negatives_yt}
\vspace{-0.3cm}
\end{table}

\subsection{Counting views in monetized view counters}
\label{sec:monetary}

Having established a baseline for the penalization of views in public view counters, with  $i)$ obviously aggressive fake view patterns, and $ii)$  real viewers, we now look  at the penalization in view counters for monetized content.

For the following set of experiments, we study the performance of audit system for monetized views of \youtube and \dailymotion. These services monetize views by serving ads to viewers and sharing revenue with their uploaders/web masters. 
We consider a view from the probe as monetized, iff  a video ad is served to it, and the probe views the whole ad and the video. We therefore count only views that we generate, and are served an ad. Then, using the reporting tools provided by each portal, we compare the number of monetized views generated by the probe and the numbers reported by the portals.

To conduct monetization experiments, we register several accounts and their associated videos in the monetization program of each portal. Since \dailymotion only monetizes videos embedded in external webpages, we create external webpages to embed the videos. The web master accounts for these pages are then associated with uploader accounts. To monetize content on \youtube, uploaders must register their channels to AdSense, Google's monetization platform, and indicate which videos to enroll.  While for \dailymotion, the probes  direct their views to the external webpages we create for experiments, for \youtube, we direct  views to the \youtube URLs for the experiment videos.

We have developed several techniques to identify whether ads are really served in a probe's views. For \youtube, we analyze the packet data of the view session, and identify ads by deciphering the ad serving protocol. While for \dailymotion, we have developed an image analysis tool, that analyzes snapshots of the view sessions, and looks for indicative signs that an ad is being served, such as the text box used to indicate the remaining time for an ad to finish playing.

Finally, we run these experiments using the default values for the probe parameters given in Table~\ref{tab:botparams}, with the number of views per video, per day set to 20, and using a single IP address per probe (from the pool of IPs in Spain and Germany). Each experiment lasts for 20 days on each portal and is repeated four times. 

From the traces we know that less than 0.04\%($\TONE$), and 0.01\% ($\TTWO$) of IPs in the traces performed more than 20 views per day to a single video. We therefore consider our experiment configuration to be aggressive, and the fake views easy to identify. Moreover, as monetized fake views translate to direct costs to advertisers, we expect both portals to be stricter in the identification of fraudulent views for monetized content.

\begin{figure}
  \centering \includegraphics[width=0.42\textwidth]{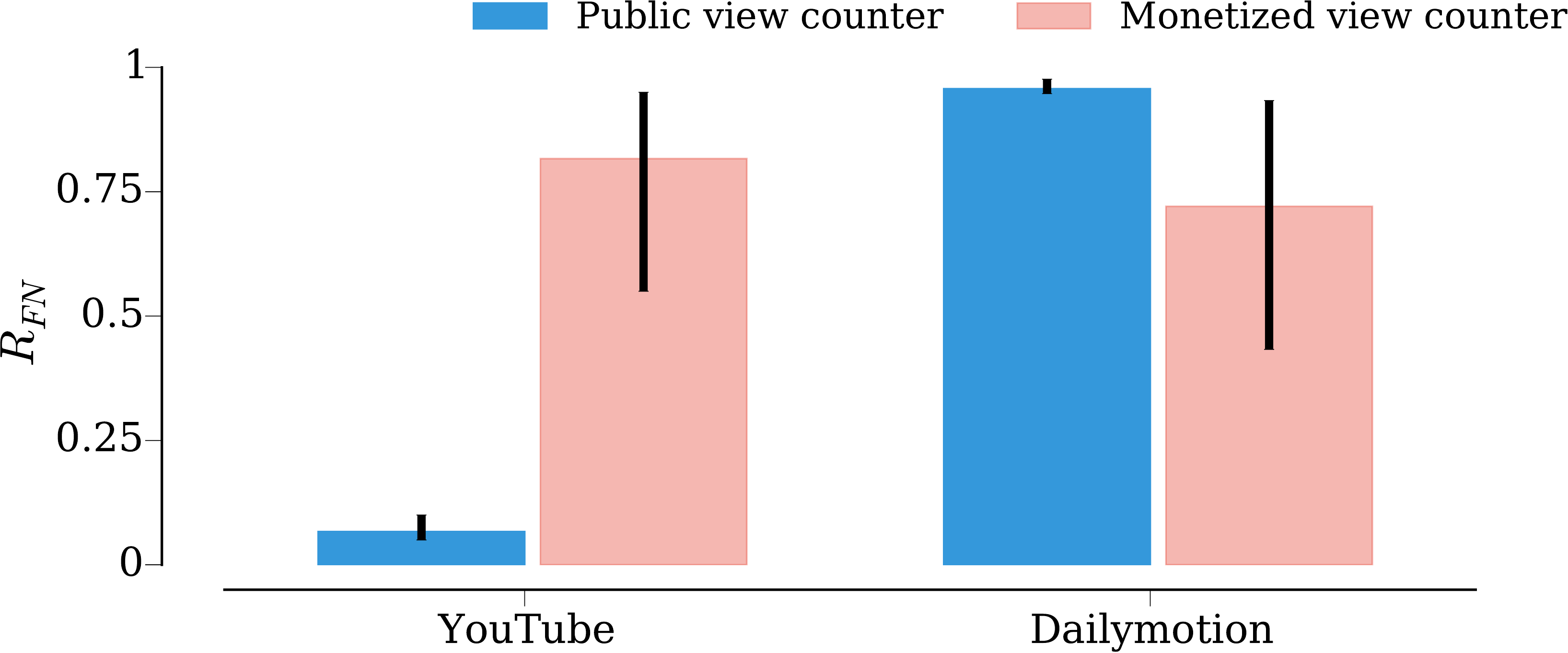} \caption{Comparison of false negative ratio for the number of views in the public and monetized view counters for \youtube and \dailymotion.}
  \label{fig:monetization-comparison}
 \vspace{-0.3cm}
\end{figure}

To evaluate the monetized view auditing systems deployed by \youtube and \dailymotion, we report their respective false negative rates ($R_{FN}$) in Figure~\ref{fig:monetization-comparison}. We compare the $R_{FN}$ in the number of views reported by public and monetized view counters. Again, the main bar depicts the average value across the experiments, and the error bars give the min-max value of the $R_{FN}$ across the experiments.

\dailymotion shows the expected behavior, and discounts a larger number of fake views from the monetization view counter (avg. $R_{FN}$ = 72\%) with respect to the public view counter (avg. $R_{FN}$ = 97\%). Despite this improvement, the view audit system for monetized views still performs poorly and roughly 3 out of 4 fake views are monetized, even under the aggressive configuration of the experiment.
Surprisingly, \youtube results are in contradiction with our expectations. We observe that \youtube's view auditing system is more permissive for monetized views (avg. $R_{FN}$ =  82\%), when compared to public view counter (avg. $R_{FN}$ = 7\%). 

This unexpected result has been reported previously by YouTube users.\footnote{see \url{https://plus.google.com/100368302890592068600/posts/1sEuu94EjuV}} \youtube support stated that discrepancies may be due to users watching the video ad, but not the video, and in that case, a view is monetized but not counted by the public counter. However, since we instruct the probes to view both the ad and the video in full, this does not hold in our case.

Another possible source of the discrepancy may be due to \youtube performing post hoc, rather than real time auditing\footnote{Post hoc auditing may be preferred since this approach obstructs reverse engineering efforts by fraudsters in comparison to real time detection.} to identify suspicious activity~\cite{Chen:2014:FVA:2597176.2578263}. 
However, more than 11 months have passed since the conclusion of these experiments, and we have not observed any changes in the statistics reported.

\emph{In summary, we find that among the online video portals studied, only \youtube deploys a sufficiently discriminative video view auditing system. However, we observe that \youtube appears to only penalize views for the public view counter.} Having observed that \youtube deploys the most discriminative view auditing system, in the rest of the paper, we extend the analysis to help understand some of the variables that it considers, for public (Sec.~\ref{sec:youtube}) and monetized (Sec.~\ref{sec:youtube_monetization}) view counters.

\section{YouTube's Audit System for Public View Counter}
\label{sec:youtube}

In this section, we explore some of the different variables that are considered by the view auditing system deployed at \youtube. Because we have adopted a black box method to testing, we focus on meaningful parameters that are easily accessible to fraudsters. As indicated earlier, bots executing attacks on \youtube manipulate a similar set of parameters~\cite{Tubrosa,YoutubeBotViews,YouTubeBotViewsProxies}. Note that in the remainder of the paper, unless otherwise stated, all described experiments are repeated 3 times using different IP addresses from our /24 IP prefixes in Spain and Germany, to detect potential geographical biases in the measurements.

\begin{figure*}[!htb]
\minipage[t]{0.325\textwidth}
\hspace{-0.5cm}\includegraphics[width=1.1\linewidth]{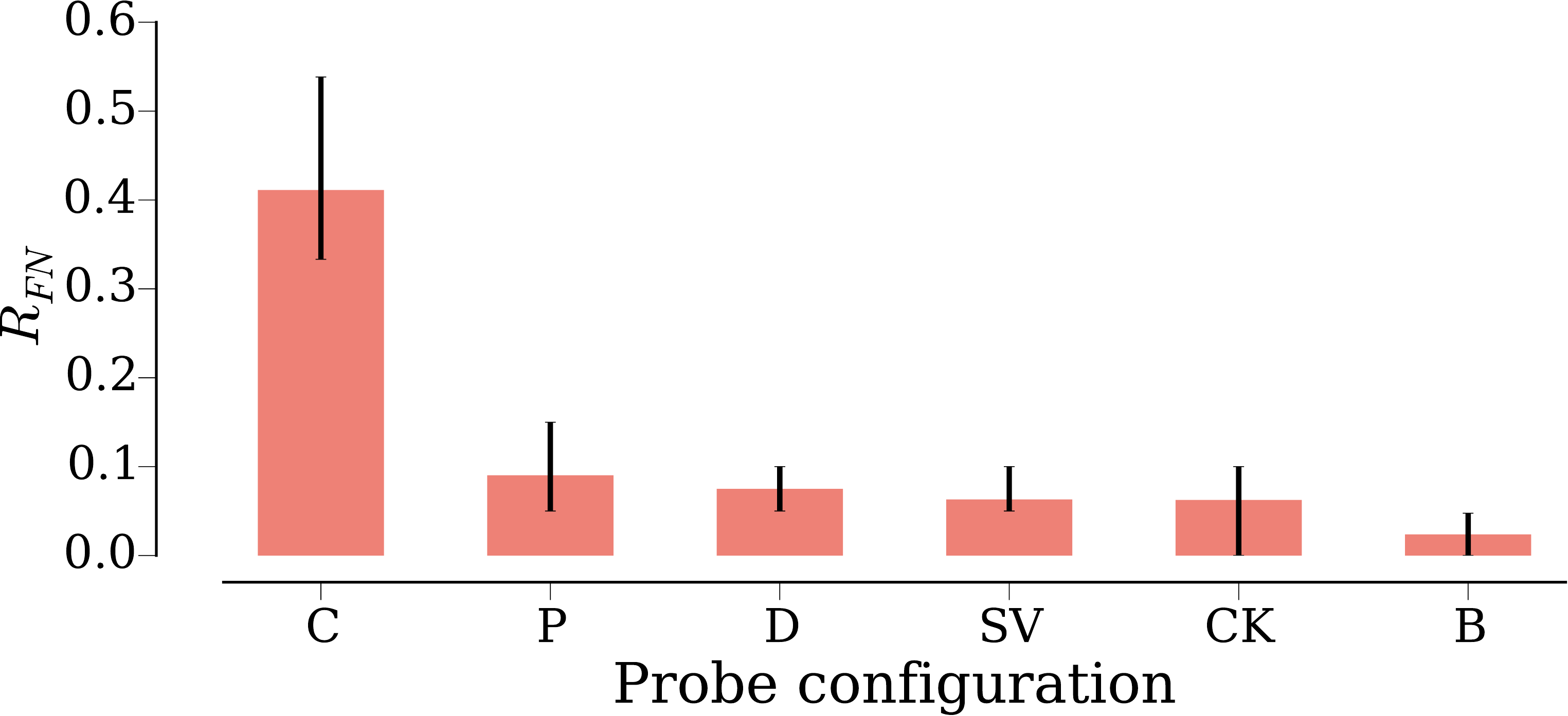}
\caption{False negative rate obtained for each of the experiment configurations.}
\label{fig:bots_comparison}
\endminipage\hfill
\minipage[t]{0.325\textwidth}
\hspace{-0.2cm}\includegraphics[width=1.1\columnwidth]{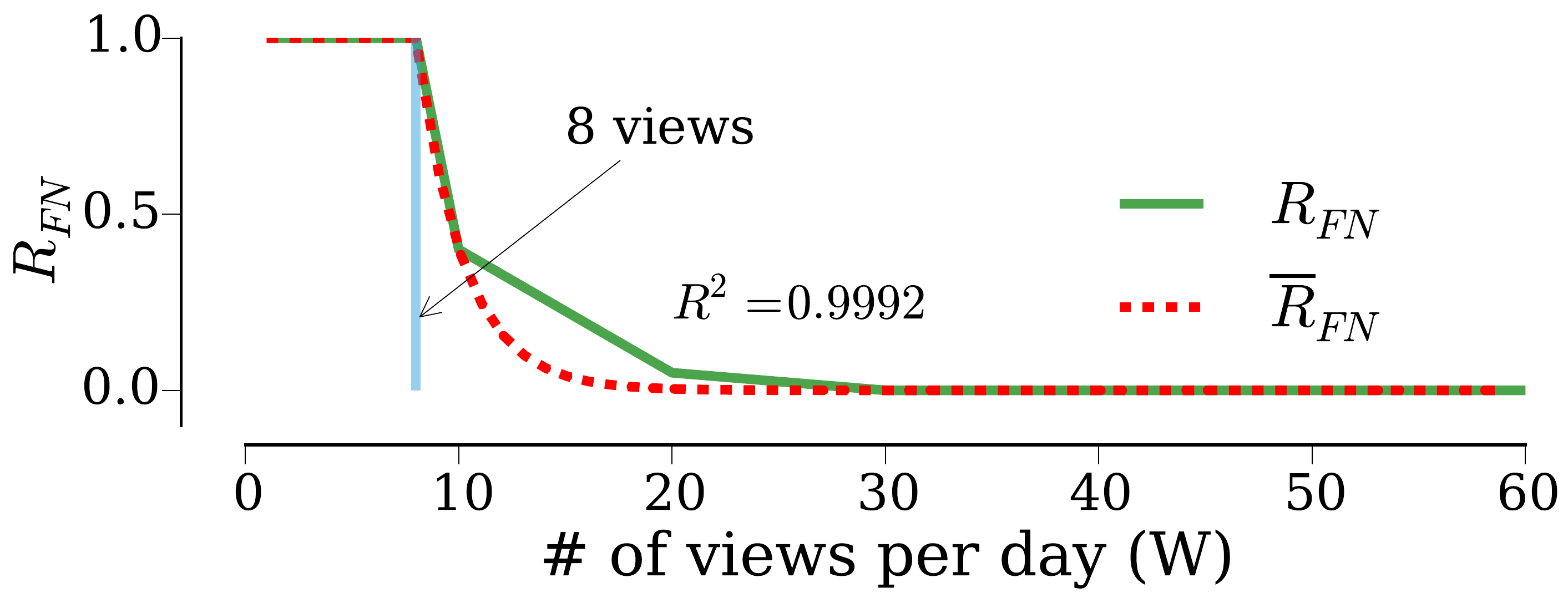}
\caption{False negative rate to one video depending on the number of views per day.}
\label{fig:punishment}
\endminipage\hfill
\minipage[t]{0.325\textwidth}
\hspace{0.2cm}\includegraphics[width=1.1\linewidth]{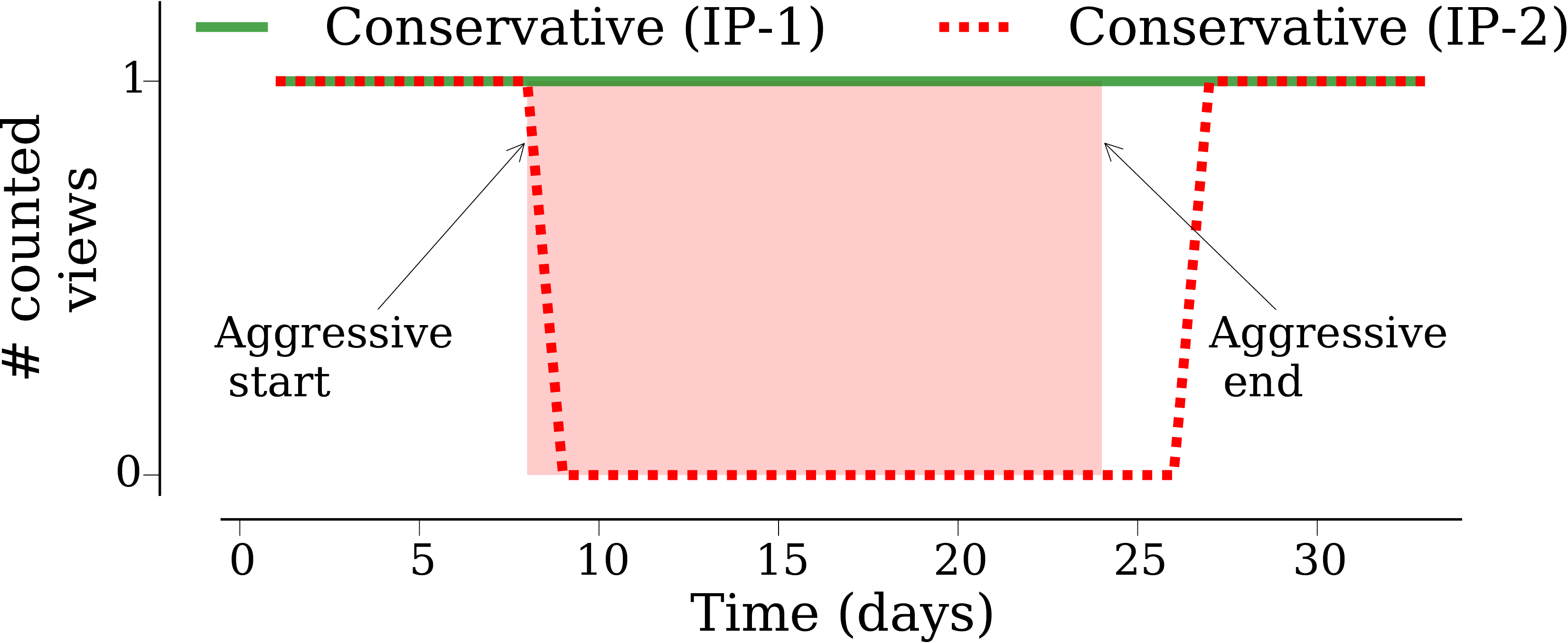}
\caption{Number of views counted by \youtube for both IP-1 (\emph{conservative} probe) and IP-2 (\emph{conservative} and \emph{aggressive} probe).}
\label{fig:ip_blacklisting}
\endminipage
\vspace{-0.3cm}
\end{figure*}

\subsection{Parameters used in the detection}\label{sec:paramsyt}
In order to explore the parameter space, and isolate their impact in the audit system, we configure the probes to run individual behaviors (configurations of the parameters listed in Table~\ref{tab:botparams}).

In the following, each probe instance uses a single public IP address chosen from the pool, and performs 20 views per day, to the same video for 8 consecutive days, and repeats this 5 times to evaluate the rate of false negative ($R_{FN}$), i.e., the ratio of fake views misclassified as valid and counted. Based on the results in Section~\ref{sec:monetary}, we expect that the probe behaviours are flagged as suspicious, and trigger the view audit system of \youtube.

Next, we describe the probe behaviors we use in the experiments. Note that unless specified, we set all parameters to the default values given in Table~\ref{tab:botparams}.

\vspace{0.1cm}
\noindent \textbf{- Deterministic (D)}: The goal of this  behavior is to define a simple, and completely deterministic pattern of views.  This behavior  eliminates any randomness  by setting to constant values the view time (40 secs.) and the time between views (72 mins.). All other parameters take their default values from Table~\ref{tab:botparams}. We expect this behavior to be easily identified.

\noindent  \textbf{- Vary view burst (B)}: The goal of this behavior is to study the impact of making views in bursts. In particular, the probes run the {\bf D}eterministic behavior, setting the time between consecutive views to 0, and generating a burst of N = 20 consecutive views every day. The time between consecutive bursts can be configured, and is set to 24 hours in the experiments. Since bursts of views from a given IP address, to a single video are atypical for users in $\TONE$ and $\TTWO$, we expect this behavior to be easily identified, and to have low false negative ratio.

\noindent  \textbf{- Vary inter-view wait time (P)}: The goal of this behavior is to measure the impact of varying the time between views over a day. The probe runs the {\bf D}eterministic behavior, but varies the time between two consecutive views.  With this behavior, we aim to determine whether adding some noise to the inter-arrival pattern of views has any impact when compared to a deterministic pattern In the following, we use a Poisson with $\lambda = 20$.

\noindent \textbf{- Short Views (SV)}: The goal of this behavior is to measure the impact of making very short views to videos. In the following, the probe runs the {\bf D}eterministic behavior, but sets the duration of video views to 1 sec. Since consecutive short views  are atypical for real users, we expect to see this behavior will be heavily penalized.

\noindent \textbf{- Cookies (CK)}: The goal of this behavior is to measure to what extent audit systems rely on user identifiers when auditing views. We use cookies since they are the most commonly used method to track users~\cite{Englehardt:2015:CGY:2736277.2741679}. We consider the extreme case in which the probe uses the {\bf D}eterministic behavior, and performs all views using the same cookie.

\noindent \textbf{- Complete (C)}: The goal of this behavior is provide a baseline by emulating some real-user like features. Therefore we enable all the parameters in Table~\ref{tab:botparams}, except the cookies. Specifically, the view duration time and wait time between views are set to Poisson processes with $\lambda$ = duration of the targeted video, and $\lambda$ = 72min,  respectively. Finally, the Referrer and User-Agent fields are selected randomly. Given the variation in the parameters, we expect this behavior to be the least penalized.

For each behavior, Figure~\ref{fig:bots_comparison} gives the average and max/min $R_{FN}$.  As expected, the {\bf C}omplete behavior yields the highest false positive rate ($\approx$40\%), and is on average 4x larger with respect to the other behaviors ($R_{FN} <$ 10\%). This indicates that adding some randomness to basic HTTP parameters such as the User-Agent, or the Referrer makes it significantly harder for \youtube to detect fake views.

Looking at the impact of varying the wait time between views ({\bf P}, {\bf D} and {\bf B}), we observe that the view audit system penalizes {\bf B}ursty behavior the most heavily, discounting  98\% of the views. Comparing the {\bf D}eterministic and the {\bf S}hort {\bf V}iews behaviors, contrary to our expectation, they are both similarly penalized. We observe that the audit system counts as valid 7\% and 6\% of views for the {\bf D} and {\bf SV} configurations respectively . Finally, we observe no significant change to enabling/disabling user tracking via the cookies. The differences in false negative ratios with cookies({\bf CK}) and without ({\bf D}, {\bf SV}, etc.) cookies are negligible.

In summary, \emph{we find that \youtube is able to identify the simplest suspicious behavior patterns, schemes using static HTTP connection parameters are easily identified. Indeed, the view audit system is able to remove more than 90\% of fake views generated under these attack configurations. We observe however that adding some variability to HTTP connection parameters may increase the effectiveness of attacks up to $\sim$30\%}. While these results explain the false negative rate difference between the considered configurations and the benchmark, they do not explain the significant number (60\%) of discounted fake views common to all the configurations. The only variable common to all the configurations, and which may be responsible for such large a penalization is that they each perform their views from a unique public IP address. This along with the fact that IP addresses are one of the strongest online users identifiers~\cite{Chen:2014:FVA:2597176.2578263}, and one of the key parameters many security online services use~\cite{Metwally:2007:DDC:1242572.1242606, Collins:2007:UUP:1298306.1298319, Ramachandran:2006:UNB:1151659.1159947} leads us to believe that the video viewing pattern from an IP address is a key element for the fake view detection mechanism of \youtube. We analyze this hypothesis in the next subsection.

\subsection{Influence of Video Viewing Pattern in the detection}\label{sec:paramsip}

In this subsection we analyze the response of \youtube's view audit systems to the fake view patterns of an IP address. We first look at the impact of view patterns to a single video, then explore the cases for a single IP viewing multiple videos, and finally a single video receiving views from multiple IP addresses.


\vspace{0.1cm}
{\noindent \bf One video, One IP address\\}
We start by examining how \youtube discounts the views generated by a single IP address to a single video. In particular, we are interested in understanding how the view penalization threshold(s) are triggered, when varying the number of views per day. We conduct a simple experiment, in which the probe generates $W = [1, 4, 7, 8, 9, 10, 20, 30, 40, 50, 60]$ views per day, to a given video, for 8 days. We use the previously defined {\bf D}eterministic behavior for this experiment.

The results of this experiment are presented in Figure~\ref{fig:punishment}, which reports the $R_{FN}$ for the different numbers of views
($W$). We observe that the view audit system counts all the views up to a rate of 8 per day. From 9 views on, the $R_{FN}$ decays exponentially and is 0 for more than 30 views per day. We observe that the $R_{FN}$ with respect to the views per day ($W$) follows an exponential decay function, and can be modeled with the following parameters, with an R$^2$ = 0.999:
\vspace{-0.1cm}

\begin{equation*}
 \overline{R_{FN}}(W)= \left\{
 \begin{array}{ll}
  1 & \mbox{if } W \le 8,\\
  e^{-0.455n} & otherwise
  \end{array}
  \right.
\end{equation*}
\vspace{-0.1cm}

For the previous experiments, we used newly uploaded videos. To understand whether this has any impact on the results obtained, we look at the response of the audit system when we generate views for videos previously uploaded to \youtube and are moderately popular, and repeat the experiment. With the permission of the uploaders, we use two videos with roughly 12K (100 in the last month) and 300K (5K in the last month) registered views at the start of the experiment. To identify the activity of the probe in the results, we configure it to use very rare User-Agents (Bada, HitTop, MeeGo and Nintendo 3DS). Before starting the experiment, we validate that the targeted videos have not received any views from the selected User-Agents in the previous 6 months using \youtube Analytics.

Setting $W = [8, 9, 10, 20]$ views per day, we find that the view audit system again starts discounting views from 8 views per day, for a given IP address, and $R_{FN}$ follows the same decay pattern. This suggests that view audit system of \youtube are triggered by a fixed threshold regardless of a video's popularity.

\vspace{0.1cm}
{\noindent \bf Multiple videos, One IP address\\}
Having observed how the views from a single IP address to a single video are penalized, we  now look at  the response of the view audit system when a single IP address spreads its views over several videos. Given the previous result, we expect that aggressive IP addresses will be heavily penalized, independent of the number of videos they target.

In the following, we first test this hypothesis, and then present the results of a large scale measurements to understand how the rate of false negative varies with respect to the number of videos viewed, and views performed, per IP address.

To begin to understand how the view audit system differentiates between IP addresses, we define two simple probe behaviors; \emph{conservative} and \emph{aggressive}. The \emph{conservative} probe performs 1 view per day, while the \emph{aggressive} probe performs 30 views per day. We set up the following experiment: in two IP addresses, IP-1 and IP-2,  we launch an instance of the \emph{conservative} probe to a different video for 34 days. Moreover, in IP-2, we also launch an instance of the \emph{aggressive} probe, starting at day 8 and stopping at day 24, while there is no aggressive probe in IP-1.

Figure~\ref{fig:ip_blacklisting} gives the number of views the audit system counts over time, for the \emph{conservative} probes in the two IP addresses. Since \emph{conservative} probes perform just 1 view per day, to the video, we expect to see either; 1, if it is counted, or 0, if it is penalized. We find that the audit system counts all the views from the \emph{conservative} probe in IP-1, and penalizes the \emph{conservative} probe in IP-2 for the days that the aggressive probe is also running from IP-2 (days 9-26). We observe that view penalization starts 24 hours after the launch of the \emph{aggressive} probe in IP-2, and ends two days after the probe stops. Repeating the experiment three times in total, we obtain the same results.  From this, we conclude that \youtube's fake view audit system labels and tracks the behavior of IP addresses based on their global view behavior across all videos that they visit.

Having observed that \youtube's view audit system labels IP addresses based on their  behavior, we now look at how it penalizes the behavior of an IP address across video views. To do so, we conduct a large scale experiment in which we perform $W = [1, 3, 5, 7, 10, 15, 20]$  views per day,  uniformly distributed across $D = [1, 3, 5, 7, 10, 15, 20]$ videos (with $W \ge D$), over a period of 7 days.\footnote{Note that we only run experiments for $W \ge D$. For instance, in the case of $W=5$ we run experiments for $D = {1, 3, 5}$.} 
In total, we ran 28 combinations of views and videos. Finally, we use the {\bf D}eterministic behavior for the probe.

\begin{figure}[t]
 \centering
 \includegraphics[width=0.40\textwidth]{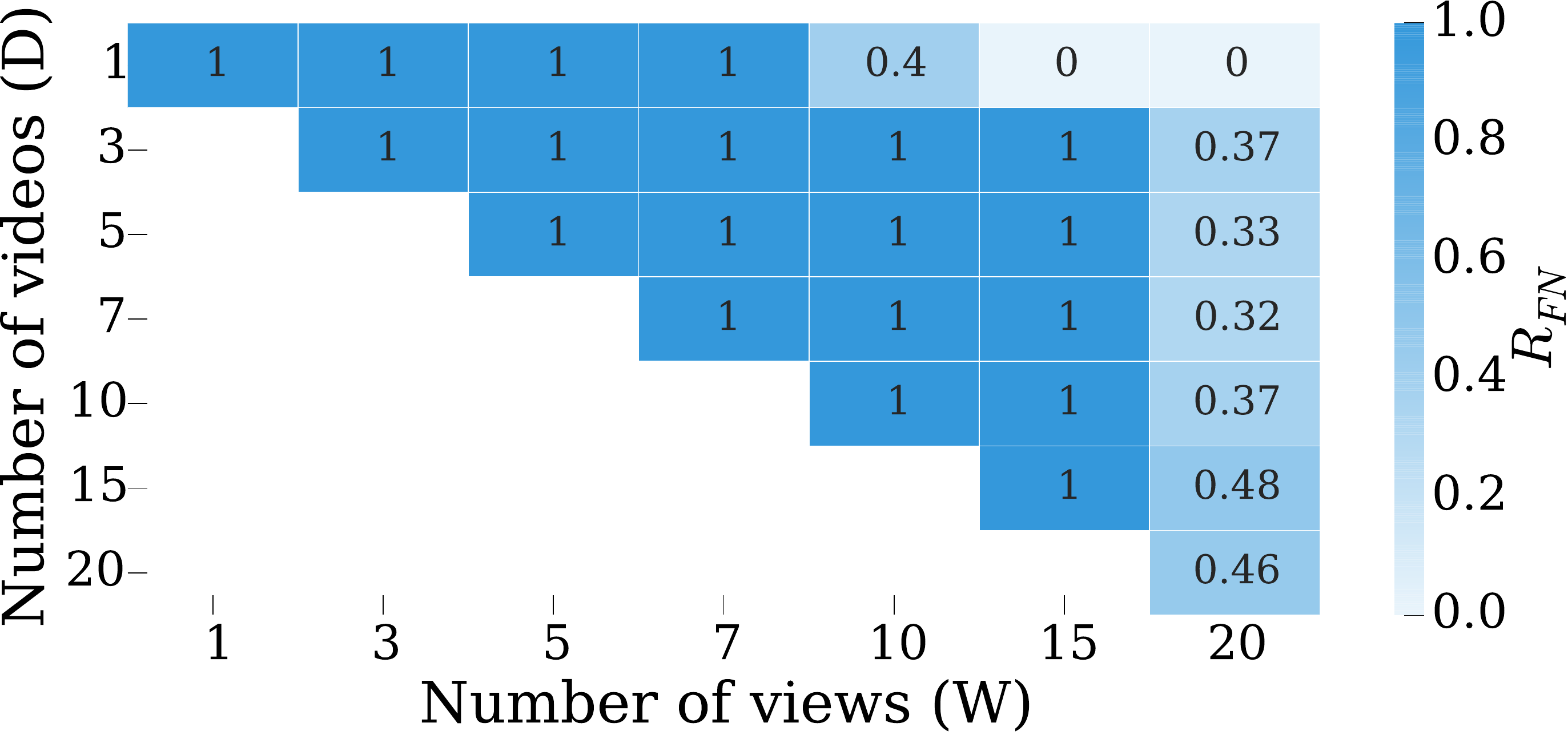}
 \caption{$R_{FN}$ for several combinations of the number of views $W$ and the number of watched videos $D$.}
 \label{fig:albert}
 \vspace{-0.3cm}
\end{figure}

Figure~\ref{fig:albert} reports the $R_{FN}$ across the 28 combinations considered. Looking at the evolution of $R_{FN}$ for a fixed number of videos, we observe the exponential decay revealed in Figure~\ref{fig:punishment}. However, in this case, view penalization is triggered after 15 views per day, when a viewer watches 3 or more distinct videos per day ($D \ge 3$), whereas in Figure~\ref{fig:punishment} it was triggered after 8 views per day (for $D=1$).
With respect to the evolution of $R_{FN}$ for a fixed number of daily views, we observe that when all views are to a single video, the penalization is much more severe, than when they spread across three or more videos. 

\vspace{0.1cm}
{\noindent \bf One video, Multiple IP addresses\\}
Having established, that an IP address is tracked across video views, we now look at the response of the view audit system, when the views to a given video are distributed across several IP addresses.
To this end, we use 70 different PlanetLab proxies, and divide them in 3 independent groups of different size $N=[10, 20, 40]$. We assign each group of proxies a different video on \youtube, and configure each proxy to generate views to its corresponding video. We again utilize the {\bf D}eterministic behavior of the probe, and report the results with each PlanetLab proxy group to generate 3 views per day.  Overall, the experiment generates 30, 60 and 120 views per day to a video, which should  result in $R_{FN} = 0$, if coming from a single IP address.

From this experiment, we observe that the growth in number of views over time is linear for all behaviors, and that overall $R_{FN} >$ 73\% in all three experiments. This indicates that distributing activity across multiple IP addresses results in a substantial increase in the $R_{FN}$ enabling attackers to inflate view counters easily.

This experiment suggests that \emph{YouTube is vulnerable to attacks that employ many IP addresses} (such as those from botnets), and such attacks can apparently achieve an arbitrarily large number of views. In fact, it is easy to find paid services that offer to inflate the view counter of \youtube (and other video platforms) videos up to tens of thousand views in a short period of time and at a low price (e.g. \cite{viewbros,qqtube,buildmyviews}).

\subsection{Impact of NATed IP addresses on the audit system}
As NAT devices aggregate traffic, they typically contain the video viewing activity from multiple, usually private, IP addresses. In large NATed networks, such as campus networks, corporate networks, and in some cases ISP networks, this activity may be significantly large.

Therefore, in the following set of experiments, we investigate how the view audit system of \youtube penalizes the views originating from NATed networks. To do so, we install the probes on three machines accessing the Internet from NATed networks located at three different locations, and we configure them to perform 20 (Location 1), 75 (Location 2), and 100 (Location 3) views per day for a period of 8 days. We again use the {\bf D}eterministic behavior. 

Table~\ref{tab:natRFP} reports the $R_{FN}$ for each experiment along with information of the different NATed scenarios. Note that, although the probe generates views aggressively, the $R_{FN}$ is surprisingly large in all cases. This suggests that the \youtube's view audit system has problems in properly identifying suspicious activity from NATed networks. To confirm this finding, we separately analyze the $R_{FN}$ on working days and days off (i.e., weekends and holidays) in Location 2, and run the experiment for 194 days. Note, during working days the volume of NATed traffic from the network is high,  whereas it is low during the days off. Figure~\ref{fig:nat} shows the distribution of the daily false negative rate for working days and days off in the boxplots. The results confirm that \youtube discounts almost all views during days off, i.e., when the traffic is more exposed, but has problems in discount views (median $R_{FN}$ = 60\%) for workdays, i.e., when the views are hidden by larger volumes of traffic.
Hence, this suggests that malicious users can dramatically increase the efficiency of their activity by gaining access to machines located behind large (active) NATed networks, e.g., a public campus network.

\begin{table}
\small
\centering
\begin{tabular}{ c || c |c |c| c}
\textbf{Experiment} & \textbf{W (views/day)} & \specialcell{\textbf{\# U (users } \\ \textbf{behind the NAT)}}  & \textbf{U/W} & $R_{FN}$ \\
\hline
Loc. 1 & 20  & $\sim$50 & $\sim2.5$ & 0.9 \\
\hline
Loc. 2 & 75  & $ \sim$100 & $\sim1.33$ & 0.43 \\
\hline
Loc. 3 & 100 & $\sim$50 & $\sim0.5$ & 0.36 \\
\end{tabular}
\caption{$R_{FN}$ and information about the three scenarios for the experiments we conduct from NATed IP addresses.}
\label{tab:natRFP}
\vspace{-0.4cm}
\end{table}

\begin{figure}[t]
\centering
 \includegraphics[width=0.42\textwidth]{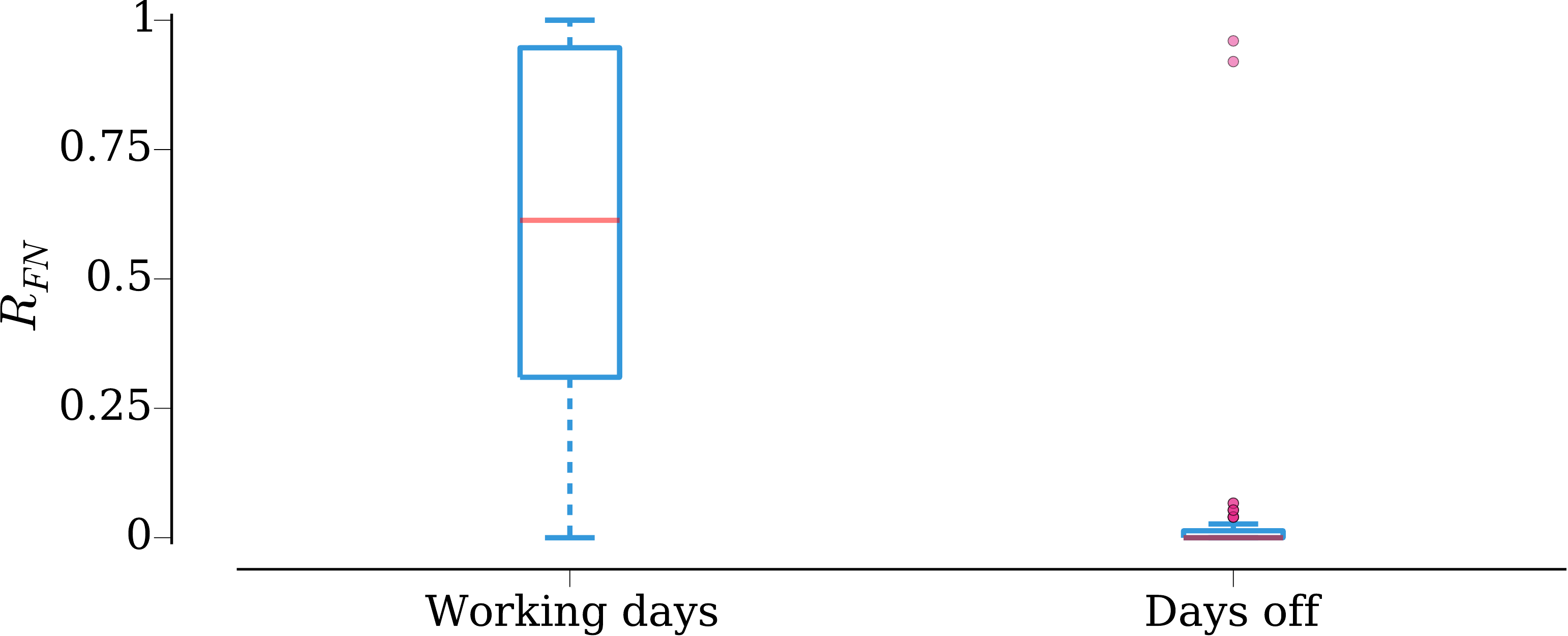}
\caption{Distribution of daily $R_{FN}$ for working days and days off for Location 2.}
\label{fig:nat}
\vspace{-0.3cm}
\end{figure}


\subsection{Punishment of IP prefixes}

Our analysis so far shows that an IP address is punished by its global
behavior. In this subsection, we go one step further to analyze whether
YouTube's fake view detection algorithm punishes ranges of IP addresses when
one of them is misbehaving. Note that punishing IP prefixes due to
the misbehavior of a single IP address is a common technique that, for
instance, we have experienced when querying BitTorrent trackers in
previous studies~\cite{Cuevas:2013:UIC:2578911.2578917}. In addition,
some existing solutions propose to consider IP address within the same
prefixes as it has been observed that botnet-infected machines choose as potential future members of
the botnet machines within the same IP prefix~\cite{Collins:2007:UUP:1298306.1298319}.

We perform a similar experiment to the one described in
Section~\ref{sec:paramsip}. We start an instance from IP address \emph{IP-A}
that behaves properly and makes 1 daily view to a video. After a few
days, we start a second instance from IP address \emph{IP-B}, which misbehaves and
performs 20 daily views to a second video. Note that \emph{IP-A} and \emph{IP-B} 
belong to the same /X prefix. We conduct this experiment
for values of X ranging between 24 and 30 and we
did not observe any punishment. Therefore, we
conclude that \youtube detection mechanism does not punish consecutive IP
addresses belonging to the same /24 onward.

\subsection{Detection Time}

The results of Figure~\ref{fig:ip_blacklisting}, as well as those of other
experiments, indicate that the punishment does not start right after
the IP address begins to misbehave. This suggests that \youtube's fake views
detection mechanism requires some time before it starts
punishing a misbehaving IP address. Our aim in this subsection is to
quantify this ``detection time'' with respect to the past history of an
IP address. In particular, we consider three types of IP addresses based on
their history: $(i)$ a fully-clean IP address that we have never used to
connect to \youtube, $(ii)$ an IP address that we have used before to watch
\youtube videos but has never shown a misbehaving watching pattern; and $(iii)$
an IP address that has shown a misbehaving watching pattern in the past. For
each one of these IP addresses, we start 7 instances of our software performing $W = 3, 5, 7, 10, 15, 20$ and
$25$ views per day, respectively. This aggressive behavior guarantees
that the fake view detection system will mark the IP addresses as suspicious and will discount their views. Our results
show that the system punishes the fully-clean IP address after 12 days,
whereas it starts punishing the other two IP addresses one day after the experiment starts. Therefore, it seems that \youtube monitors and logs any IP address that connects to the system, and as soon as an already logged IP address misbehaves, the
\youtube detection mechanism start discounting its views just after one day. However, for
IP addresses which are unknown to the system, the detection mechanism is much more
conservative and does not discounts their views until some days have passed.

\vspace{0.1cm}
In summary, \emph{the view audit system of \youtube implements an exponential discount factor of the number of views performed from a single IP address that increases with the rate of views. However, the results show that some simple  modifications in a fraudster's strategy can considerably increase the false negative rate. In practice, i) adding some randomness in the HTTP connection attributes such as the User-Agent or the Referrer, ii) distributing the malicious activity across multiple IP addresses, or iii) performing fake views from NATed networks, are shown to be effective.}
\section{YouTube's Audit System for monetized views}
\label{sec:youtube_monetization}

Surprisingly, the results in Section~\ref{sec:comparison_portals} indicate that \youtube monetizes (almost) all the fake views we generate, while discounting them from the public view counters. In this section we study in more detail the audit mechanism applied to monetized views, to further understand this seemingly anomalous behavior.

\begin{figure}
\centering
\includegraphics[width=0.42\textwidth]{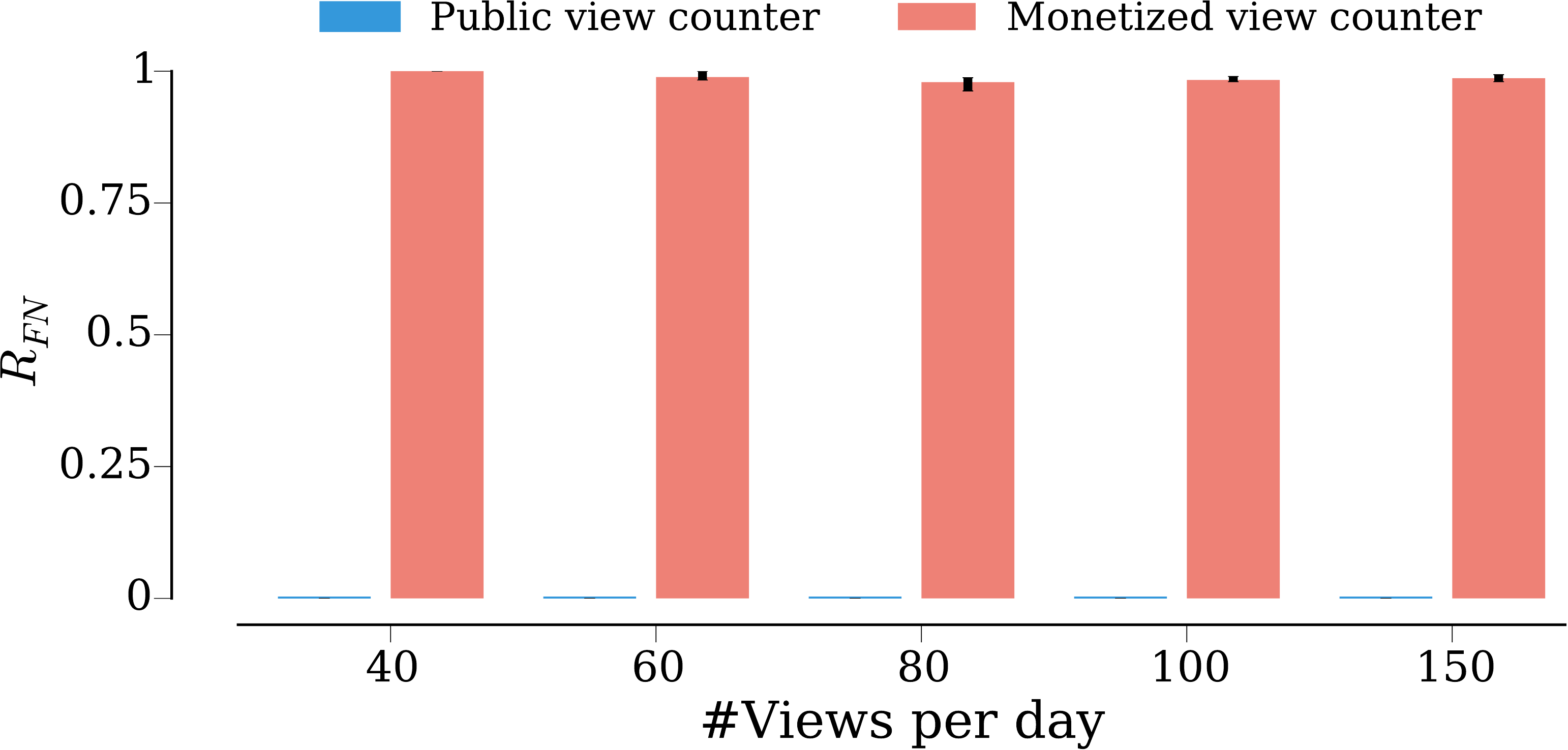}
\caption{Comparison of false negative ratio for the public and monetized view counters of \youtube for different daily rates of generated views $W$.}
\label{fig:monetization}
\vspace{-0.3cm}
\end{figure}

We reuse the configuration described in Section~\ref{sec:monetary} for \youtube, and conduct a new set of experiments, whereby we increment the number of views per day the probe generates from a single IP address, to a single video. In particular, we set $W=[40, 60, 80, 100, 150]$ to cover a wide range of aggressive configurations. We conduct each experiment for 10 days. Figure~\ref{fig:monetization} reports the $R_{FN}$ for both the public and the monetized view counters. Again, the main bars and error bars represent the average and the max/min $R_{FN}$, respectively.

We find that the monetized view counter's audit system penalizes a negligible portion of views in all the considered configurations, 
 while the public view counter's audit system penalizes most of the fake views. These results confirm the preliminary observation in Section~\ref{sec:comparison_portals}; \youtube applies different penalization schemes to the fake views in the monetized  and public view counter, with the former being much more permissive than the latter.

\subsection{Counting monetized views from the advertiser's perspective}

To gain insight into the monetary implications of the above finding, we designed a tailored experiment in which we assume the role of an advertiser exposed to fraudulent views. To do so, we first create an advertiser account using the Google AdWords service. AdWords enables us to configure advertising campaigns in \youtube, so that our video ads can target \youtube videos whose uploaders participate in the monetization programme.  We then create a video ad and build an advertising campaign to target \underline{experiment} videos that we have previously uploaded to \youtube. In this way, we play both the role of the advertiser and the publisher in the campaigns, and can build a complete picture of the trade.

AdWords offers a wide range of tools to aid in the design of video advertising campaigns. Advertisers can tailor campaigns to reach specific \youtube viewer demographics (per interests, country, language, gender, age), or target specific \youtube videos. With the aim of checking if \youtube actually charges advertisers in presence of fake views,
we configure a campaign to target the views from the countries where the proxies are located (accepting all languages, genders and ages) and headed to the experiment videos. Then, we use the probe to generate views to these videos.

\youtube deploys a sophisticated bidding algorithm that selects in real time the ad to target to a specific video. Briefly, this algorithms implements a variant of a Vickrey auction, named Generalized Second-Price auction~\cite{Edelman05internetadvertising} for which the  winner (advertiser) pays the price of the second highest bid. Note that winning bids vary over time and targeted videos. In addition to the bid price, the algorithm also considers other parameters including the profile of the viewer watching the video, the advertiser's daily budget, etc.

In setting up these experiments, we faced several challenges to configure a successful campaign able to target a large number of ad views in the videos. Our initial trails were unsuccessful; we used a small daily budget of 50\euro~and the campaign had an unusual configuration, since it targeted very specific and relatively unpopular videos. To overcome this,  we took advice from an online advertising expert to; $i)$ increase the bidding prices per ad view up to 10-15\euro~ (the recommended bid price for \youtube was 0.04-0.05\euro), $ii)$ configure the video uploader's AdSense account to accept only the specific type of ads defined in the campaign, $iii)$ configure different campaigns with different accounts that compete for the same videos (viewers), $iv)$ to vary the pattern of views to the videos.

Having done so, we launched new experiments, whereby we targeted a set of videos from different IP addresses and different rates of views per IP address (between 10 and 70 views per IP address). In particular, the campaigns targeted 14 videos, using the {\bf D}eterministic configuration of the probe. 
Of the 14 trails, 5 videos were able to attract ad views from the campaigns, meaning that we bid and won - in effect the ads were targeted to the uploaded videos, and watched by the probes, which are configured to view in full any ad target, as well as the video.

Table~\ref{tab:view-patterns} summarizes the main characteristics of the view pattern configuration of these videos. Moreover, it shows the number of monetized ad views from the campaigns, as well as the number of counted views in the public view counter for the days of the experiment in which our ad was delivered.
We observe that in all the cases the number of monetized views are larger than the number of counted views, i.e., views considered suspicious are removed from the public view counter, but monetized.

Our videos received a total of 301 ad views in 5 days.\footnote{
Note that after finishing the experiments, these videos have received only 16 views in 8 months. Based on this, we have high certainty that our video ad was only viewed by the probes, and not by legitimate users.}
In the case of Video 3 and 4, views were initially added to the bill of our advertiser account. However, 5 days after the first ad view was delivered, \youtube rightfully labeled the probe's activity as suspicious and suspended the video uploader account in AdSense. In addition, \youtube notified us via email of the suspension of the uploader's account due to suspicious activity . Finally, the ad views associated to fake views were removed from the advertiser account and 4.85\euro ~refunded. We believe that the peculiar setup of the campaign, coupled with the aggressiveness of the experiment triggered  alarms in \youtube's view audit system. In case of Video 4, we repeated this experiment twice obtaining the same result (AdSense account closed).

In the case of Video 1 and 2, 91 ad views were shown, for which we were charged 5.65\euro, whereas just 25 views were counted in the public view counter.
Google indicates through its AdWords support website that \textit{"If we find invalid clicks that have somehow escaped the automated detection in the past two months, we'll give you credit for these clicks"}~\cite{invalidclicksyoutube}. In the case of Video 1 and 2, all the ad views were made more than 8 months before the conclusion of this work. Therefore we can consider that the probe's actions have gone unnoticed by Google's fraud detection algorithm.


In summary, we conclude that \emph{\youtube uses a seemingly permissive view audit system to discount fake monetized views. This exposes advertisers to the risk of building their advertisement campaigns on unreliable statistics, and may make them initially burden the risk of fraud. Conversely, the public view counter is much more discriminative, demonstrating that \youtube has effective means to identify fake views. Our results also reveal that whenever the permissive threshold for the detection of fake monetized views is crossed, \youtube severely penalizes the uploader of the video by suspending her AdSense account, preventing the uploader from monetizing any of the videos associated to the suspended account}.

\begin{table}[t!]
\small
\centering
\begin{tabular}{c || c |c |c|c}
& \textbf{\# IPs} &  \specialcell{\textbf{Daily Views} \\ \textbf{per IP}} &  \specialcell{\textbf{Monetized view} \\ \textbf{counter}} &  \specialcell{\textbf{Public view} \\ \textbf{counter}}   \cr\hline
Video 1    &  1	& 10     & 31   & 18	\cr  \hline
Video 2    &  1	& 20    & 60     & 7	\cr \hline
Video 3    &  8	& 10  & 178  &  147		\cr \hline
Video 4   &  2  &  70 & 15 (17) & 0 \cr
\end{tabular}
\caption{Experiments configuration of videos attracting ads from our advertising campaigns. The reported numbers of monetized and public counted views correspond to the sum of views of the days in which ads were shown. The number 17 for Video 4, reflects the second trail of the experiment.}
\label{tab:view-patterns}
\vspace{-0.2cm}
\end{table}

\section{Related Work}
\label{sec:relatedwork}

The research community has devoted an important amount of effort to the identification of malicious behaviors in online services and to
the design of countermeasures to such behaviors~\cite{Soldo:2012:OSF:2317330.2317335, 4509894, Venkataraman_trackingdynamic}. Similarly to YouTube's fake view detection mechanism, most of the detection system designs rely on the IP address as the main id to track and identify malicious behaviors. Some examples of such mechanisms are the classical monitoring tools looking for sources of attacks, such as port scanning~\cite{Staniford:2002:PAD:597917.597922} and DDoS attacks~\cite{PengLR04}, or the detection systems which counteract malicious users in P2P applications~\cite{Cuevas:2014:TSS:2580131.2580678}. Only those systems requiring the user registration to gain access to the service, e.g., Online Social Networks, implement detection mechanisms that use both the IP address and the user id as basic units to detect inappropriate behaviors. For instance, Facebook traces the requests pattern from a given account, if it is unusual, the user is warned and if the behavior persists the account is closed~\cite{6027868}. 

More recently, the rapid proliferation of botnets and specialized bots to attack specific services has led the research community to work on the identification, characterization and elimination of botnets and bots~\cite{Karasaridis:2007:WBD:1323128.1323135, Xie:2008:SBS:1402946.1402979, Lee10uncoveringsocial,6682691,Thonnard:2011:SAS:2030376.2030395,5665793,conf/uss/StringhiniHSKV11,Stringhini:2014:HBS:2590296.2590302,Bilge:2012:DDB:2420950.2420969}. 
Additionally, following a similar methodology to the one we use in this paper, Boshmaf et al.~\cite{Boshmaf:2011:SNB:2076732.2076746} and Bilge et al.~\cite{Bilge:2009:YCB:1526709.1526784} have developed their own automatic software to evaluate the effectiveness of the defenses of different social networks from different types of attacks such as user impersonation.

In the field of fraud detection and mitigation in online advertising, most of the literature focuses on traditional type of ads such as search or display ads. In this case, the fraud problem is referred to as ``click fraud'' since the fraudulent activity is associated to fake clicks on ads, typically performed from bots. Metwally et al.~\cite{Metwally:2007:DDC:1242572.1242606} present an early study in which they use the IP address as the parameter to detect coalition of fraudulent users or \emph{fraudsters}. In a more recent work, Li et al.~\cite{Li:2012:KYE:2382196.2382267} propose to analyze the paths of ad's redirects and the nodes found in the content delivery path to identify malicious advertising activities. Furthermore, Stone-Gross et al.~\cite{Stone-Gross:2011:UFA:2068816.2068843} managed to get access to a command-and-control botnet used for ad fraud in which the bot master sends commands with fake referrers. On a  complementary work, Miller et al. \cite{Miller:2011:WCT:2026647.2026661} study the behavior of two clicking robots: Fiesta and 7cy. 
Fiesta uses a middleman that probably shares its revenue with advertiser sub-syndicates. 7cy tries to emulate a human behavior and presents different behaviors depending on the location of the infected computer. Moreover, Dave et al.~\cite{Dave:2012:MFC:2377677.2377715} design an algorithm to identify click fraud from the advertiser perspective; to design this algorithm, the authors propose to measure different aspects of the user behavior in the advertiser webpage such as the mouse movements or the time spent in the website. Based on their initial work, the same authors propose, implement and test ViceROI \cite{Dave:2013:VCC:2508859.2516688}, a solution to discount fake clicks from ad networks. The basis of ViceROI detection algorithm is the fact that click-spammers will lead to a higher ROI (Return of Investment) than a legitimate publisher, as the authors claim that a realistic ROI is difficult to obtain with robots. Fraudsters can perform other types of attacks in the online advertisment ecosystem. For example, Snyder et al. \cite{weis2015} present a study of the prevalence of fraud in affiliate marketing networks. These networks encourage publishers to promote online shops on their webpages, receiving later some amount of money if the user, that has clicked in the promoted link, makes a purchase in that online shop. Fraudsters setup a webpage forcing user's browser to click the promoted link. Later if that user buys an item in the promoted online shop, the fraudster will receive credit for it. Another example is presented by Thomas et al.~\cite{43346}. They study the impact of \emph{ad injection} in the advertisement ecosystem. They identify mainly Chrome extensions and Windows binaries responsible of this source of fraud. Finally, Meng et al.~\cite{Meng:2014:YOI:2660267.2687258} present a new type of attack taking advantage of the different prices paid depending on the user's profile. They claimed that fraudsters could increase their revenue as much as 33\% by ``polluting'' user's profiles with high paying preferences. 

All the above works establish a very solid basis for the design of tools to mitigate fraud associated to traditional ads. However, they
are (in general) not applicable to fraud associated to video ads due to the different nature of video ads and click-based ads. To the best of the authors knowledge, there is only a very recent study that analyzes fraud in video ads~\cite{chen2015analysis}. 
The authors of this study use traces from a video platform in China to identify statistically outlying video viewing patterns and, based on these observations, suggest a fake view detection algorithm built on parameters such as the number of views made from an IP address to a video or the number of different IP addresses watching a given video. Unfortunately, as the authors acknowledge, they do not count with a ground truth dataset to validate their designed solution as legitimate views cannot be distinguished from fake ones in their dataset. In contrast to this work, our study focuses on five major video portals, including \youtube, the most important video platform worldwide, and pursues a different goal.
We propose a methodology to generate ground truth scenarios so that we can evaluate the performance (and unveil basic functionality principles) of different video portals' audit systems for both the number of counted and monetized views. As our methodology is extensible to other video platforms, the authors from~\cite{chen2015analysis} 
could use it to validate their proposed solution in their considered video platform. 

\section{Ethical Aspects and Feedback from the Industry}
\label{sec:ethical}
While, to the best of our knowledge, there is not a methodology that could obtain the results presented in the paper without any effect on advertisers and/or video portals, we would like to highlight that the experiments performed in this paper have an extremely low impact on both video portals and advertisers. 

Video portals have to dedicate storage resources to host our videos and bandwidth to serve views to the probes. However, the number of videos uploaded and views generated in the experiments is very small (negligible in comparison with the volume managed by these portals) and therefore has practically no impact on the operation of the services. 

Some advertisers have lost money during the experiments by having their ads shown in the videos viewed by the probes. However, based on the reported revenue by Google AdSense accounts associated to the videos, we can confirm that the total monetary losses produced by our experiments for advertisers are estimated to be lower than 6\euro. These losses are distributed across all those advertisers having their ads exposed in the videos, and thus the individual economical impact on each of them is negligible. 

In addition, we would like to highlight that we have not received any payments while running these experiments, and all the statistics we report, were retrieved from the YouTube Analytics channel page, Google AdWords page and the Dailymotion Publisher page.

Finally, we have reported our findings to \youtube and \dailymotion. \youtube has contacted us via email, stating that they recognize the validity of our results, and have not indicated any ethical concerns with our methodology. We plan to present the \dailymotion feedback and explanations, once we receive them.  Advertisers have also reacted positively to our research after the technical report of this work attracted media attention, and was published by several organizations, including the Financial Times~\cite{ft_1}, The Guardian~\cite{theguardian}, Business Insider~\cite{businessinsider} or the BBC~\cite{bbc1}. Major advertising companies and associations have welcomed the work, without raising any ethical concerns. Based on our results, they have urged Google and other major players to increase their transparency, when accounting for advertising expenditure, as well as to more effectively address the problem of fraud in online advertising~\cite{ft_2,isba}. 

\section{Conclusions and Future Work} 
\label{sec:conclusions}

To the best of our knowledge, this work is the first one to propose a set of tools to monitor and audit the view audit systems of online video portals, and enable independent and external parties to measure their performance. The application of the tools and methodology to the view counting behavior of five different video portals has highlighted some interesting observations. We find that only \youtube deploys a sufficiently discriminative view audit systems for the public view counter. All the other portals studied are susceptible to very na\"{\i}ve view inflation attacks. Clearly, this raises a problem for users with regard to the accuracy of the numbers that are reported by these portals.

A more careful analysis of \youtube's view audit systems has revealed that it is susceptible to attacks that introduce some randomness to the viewer behavior, including the use of multiple User-Agents, Referrers, multiple IP addresses, or machines within a large NATed network. These are traits that a knowledgeable attacker would be able to configure easily, and we have been reported to be common in large scale attacks using botnets. \youtube is consistently more permissive in the counts for monetized views, when compared to the public view counters. Specifically, fake views are penalized and not counted by the public view counter, but can still be monetized, i.e., have paid for ads delivered in them, and counted in the video owner's monetized views.  While \youtube is shown to strive to protect its users and clients, for example by reacting quickly when suspicious behavior is identified, we speculate that its setup seems to place an unnecessary burden of risk on advertisers. For example, fake views can be discounted equally for public and monetized counters, but they are not.

Finally, our analysis in this paper reinforce the call by industry for $(i)$ consistent and independently measurable principles on how [Supply sources (SSPs/exchanges, ad networks, and publishers)] should identify and expunge fraudulent traffic and $(ii)$ more efficient antifraud mechanisms. In future work, we intend to refine and better scale the tools, and methods developed here, and explore how to make them available to the wider community.

{
  \bibliographystyle{ieeetr}
  \small
  \bibliography{biblio}
}

\end{sloppypar}
\end{document}